\documentclass[notitlepage,10pt,aps,pre]{revtex4-1}
\usepackage[USenglish]{babel}
\usepackage[T1]{fontenc}
\usepackage[latin1]{inputenc}
\usepackage{amsmath}
\usepackage[caption=false]{subfig}
\usepackage{microtype}
\usepackage{xspace}
\usepackage{graphicx} 
\usepackage[nice]{units}
\usepackage[colorlinks=true,linkcolor=blue,citecolor=blue]{hyperref}

\DeclareMathOperator{\sgn}{sgn}
\DeclareMathOperator{\sech}{sech}

\begin{document}
\title{Line tension and reduction of apparent contact angle associated with electric double layers}
\author{Aaron D\"orr and Steffen Hardt}
\affiliation{Institute for Nano- and Microfluidics, Center of Smart Interfaces, Technische Universit\"at Darmstadt, Alarich-Weiss-Stra\ss e 10, 64287 Darmstadt, Germany}
\date{\today}

\begin{abstract}
The line tension of an electrolyte wetting a non-polar substrate is computed analytically and numerically. The results show that, depending on the value of the apparent contact angle, positive or negative line tension values may be obtained. Furthermore, a significant difference between Young's contact angle and the apparent contact angle measured several Debye lengths remote from the three-phase contact line occurs. When applying the results to water wetting highly charged surfaces, line tension values of the same order of magnitude as found in recent experiments can be achieved. Therefore, the theory presented may contribute to the understanding of line tension measurements and points to the importance of the electrostatic line tension. Being strongly dependent on the interfacial charge density, electrostatic line tension is found to be tunable via the pH value of the involved electrolyte. As a practical consequence, the stability of nanoparticles adsorbed at fluid-fluid interfaces is predicted to be dependent on the pH value. The theory is suited for future incorporation of effects due to surfactants where even larger line tension values can be expected.
\end{abstract}
\maketitle
\def\dTCL{\delta_{\mathrm{TCL}}}
\newcommand{\dG}[1]{\delta G_{#1}}
\newcommand{\dGtilde}[1]{\delta \tilde{G}^{( #1 )}}
\newcommand{\definiert}{\mathrel{\mathop:}=}
\newcommand{\definiertb}{=\mathrel{\mathop:}}
\newcommand{\vek}[1]{\mathbf{#1}}
\def\vekn{\vek n}
\newcommand{\phiord}[2]{\phi^{(#1)#2}}
\newcommand{\Etord}[2]{E_{t1}^{(#1)#2}}
\newcommand{\hord}[2]{h^{(#1)#2}}
\newcommand{\phiordunten}[1]{\phi_{12}^{(#1)}}
\newcommand{\anderstelle}[2]{\left.#1\right|_{(#2)}}
\newcommand{\partiell}[2]{\frac{\partial #1}{\partial #2}}
\newcommand{\gleich}[1]{Eq.~\eqref{#1}\xspace}
\newcommand{\euler}[1]{\mathrm{e}^{#1}}
\section{Introduction}
Since the introduction of line tension by Gibbs~\cite{Gibbs1906} as the excess free energy of any region where multiple material interfaces meet, both experimental and theoretical determination of line tension values for three-phase systems have been the subject of various studies (cf. reviews~\cite{Amirfazli2004,Drelich1996}). However, given the wide variety of reported values ranging from~\unit[$10^{-6}$]{N} to~\unit[$10^{-13}$]{N} with either sign~\cite{Amirfazli2004,Heim2013,Liu2013,Guillemot2012,Weijs2011,Berg2010,Takata2005,Takata2008,Hoorfar2005}, further research is required in order to clarify the physics underlying line tension, also regarding the large number of situations in which line tension may be important. Generally, this is the case for systems with high contact line curvature, e.g. nanoparticles at fluid-fluid interfaces~\cite{McBride2012,Cheung2009,Lehle2008,Bresme2007,Faraudo2003,Aveyard1996}, wetting of micro- and nanostructured as well as nanoporous surfaces~\cite{Raspal2012,Hoorfar2005}, or silicon nanocrystal growth~\cite{Nebol'sin2008}.\par
To further motivate this study, it is instructive to ask for the origin of the uncertainties mentioned above. First of all, line tension can only be indirectly accessed experimentally and is also strongly influenced by a number of unknown system parameters, e.g. roughness or impurities~\cite{Amirfazli2004}. Secondly, the theoretical calculation of line tension appears to be straightforward once a clear definition has been given. However, the incorporation of long-range van-der-Waals interactions into so-called local models~\cite{Bauer1999} turns out to be challenging because the choice of the a priori unknown functional form of the disjoining pressure influences the results~\cite{deGennes1985, Bauer1999, Getta1998, Solomentsev1999, Yeh1999a, Yeh1999b, Wu2004, Gomba2009, Boinovich2011}. On the other hand, density functional theory studies are usually limited to Lennard-Jones fluids~\cite{Weijs2011, Ruckenstein2010, Bauer1999}. The situation is quite different for long-range interactions due to electric double layers (EDLs) where the required expressions can be derived from Poisson-Boltzmann theory, which is the purpose of the present work. Owing to the intricate nature of van-der-Waals and other interactions contributing to line tension, here we confine ourselves solely to the electrostatic contribution.
\section{Definition and modeling of electrostatic line tension}
\begin{figure}[ht]%
\centering%
\includegraphics[width=8.65cm]{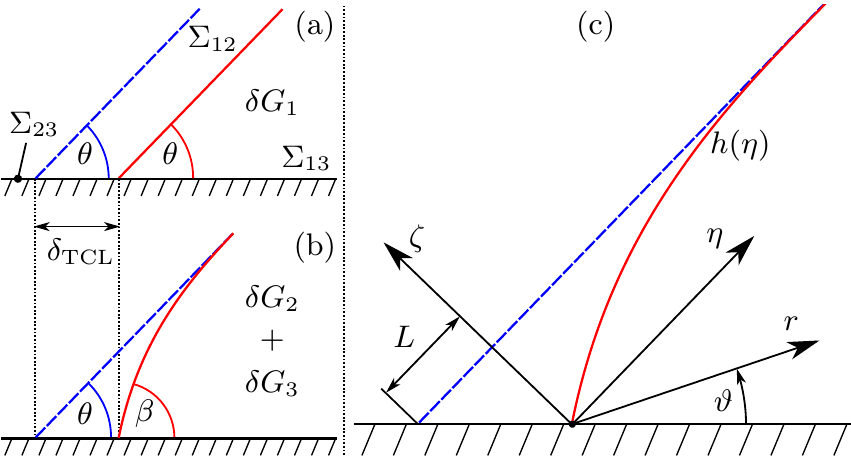}
\caption{Schematic of an electrolyte wetting a solid substrate with a macroscopic contact angle of~$\theta$. Microscopic (solid line) and macroscopic (dashed line) states with corresponding line tension contributions.}%
\label{fig:Ksys}%
\end{figure}

Similar to other works~\cite{Doerr2012,Kang2003a,Kang2003b,Das2013}, we consider a wedge-shaped three-phase contact region between a 1-1 electrolyte, a non-polar solid substrate, and a non-polar fluid. The fluid-fluid interface~$\Sigma_{12}$ (cf. Fig.~\ref{fig:Ksys}a) as well as the interface~$\Sigma_{23}$ separating the non-polar fluid from the substrate are assumed to be uncharged. Electric-double layer effects are modeled based on Poisson-Boltzmann theory. We define the electrostatic line tension as the free energy difference per unit length of contact line between a "microscopic" state, corresponding to the true equilibrium state with a deformed fluid-fluid interface, and a "macroscopic" state given by the linear extrapolation of the fluid-fluid interface down to the solid substrate (cf. Fig.~\ref{fig:Ksys}).
%
This definition of line tension is consistent with schemes to determine the molecular line tension~\cite{Solomentsev1999, Getta1998, Bauer1999, Ruckenstein2010}. In that sense, the electrostatic line tension results from the fluid-fluid interfacial deformation caused by osmotic and electrostatic forces. We remark that our model is fundamentally different from the one presented by Kang et al.~\cite{Kang2003b}, for the latter does not include interfacial deformation, which is our original motivation for introducing a line tension, and neglects the chemical contribution to the free energy. Within Poisson-Boltzmann theory\cite{Verwey1948,Overbeek1990}, the total free energy of double layer formation consists of a chemical part, resulting from the electrochemical potential difference gained by ions forming the interfacial charge, and an electric part due to electric work performed. The chemical part of the EDL free energy has to be considered along with the electric part whenever interfacial charge is changed, whereas at fixed interfacial charge only the electric part is modified. The line tension discussed in this work contains chemical and electric contributions, but is occasionally termed "electrostatic line tension" to distinguish it from the more frequently considered "molecular line tension" in the absence of an EDL. Consideration of interfacial deformation gives rise to several contributions to the free energy difference relative to the macroscopic picture (Fig.~\ref{fig:Ksys}). First, the substrate area wetted by the electrolyte is reduced, implying a change~$\dG{1}$ in the molecular and electrochemical free energy. Secondly, the area of the fluid-fluid interface is changed, leading to another interfacial energy contribution~$\dG{2}$. Thirdly, the free energy of the EDL is modified due to the deformation at fixed three-phase contact line (TCL) adding the quantity~$\dG{3}$ to the EDL contribution already included in~$\dG{1}$. Summing up these three contributions we arrive at the line tension~$\tau\definiert\dG{1}+\dG{2}+\dG{3}$.
\subsection{Electrostatic potential and interfacial shape}
Throughout the paper we use dimensionless quantities choosing the Debye length~$\kappa^{-1}$ as the length scale, the product of fluid-fluid interfacial tension~$\sigma_{12}$ and Debye length as the scale for free energies per unit length, and $\varrho_{13}/(\varepsilon_1\kappa)$, where~$\varrho_{13}$ is the interfacial charge density at the substrate-electrolyte interface~$\Sigma_{13}$ and~$\varepsilon_1$ is the permittivity of the electrolyte, as the electrostatic potential scale, respectively.\par
In what follows we assume that the non-polar fluid has a negligible permittivity as compared to the electrolyte. As a consequence, the pressure deforming the fluid-fluid interface can be expressed by the electrostatic potential~$\phi$ on the interface and the tangential electric field~$E_{t}$ in the electrolyte phase right at the interface. The interfacial shape may be described in the $(\eta,\zeta)$-coordinate system, depicted in Fig.~\ref{fig:Ksys}c, by a parameterization~$h(\eta)$ with~$h(0)=0$ and~$h'(\eta\to\infty)=0$. Defining the non-dimensional products~$\chi\definiert\varrho_{13}e/(\varepsilon_1\kappa kT)$and~$\psi\definiert\varrho_{13}^2/(2\sigma_{12}\varepsilon_1\kappa)$, where~$e$ denotes the elementary charge, $k$ Boltzmann's constant and~$T$ the temperature, we write the interfacial stress balance~\cite{Wang2011} in the form
\begin{equation}
\begin{split}
\frac{h''(\eta)}{\left[1+h'^2(\eta)\right]^{3/2}}=-2\frac{\psi}{\chi^2}\left[\cosh(\chi\phi)-1\right]_{(\eta,h(\eta))}-\psi \left.E_{t}^2\right|_{(\eta,h(\eta))}.
\end{split}
\label{eq:TPBsprungdimlos}
\end{equation}
While the left-hand side of \gleich{eq:TPBsprungdimlos} represents the interfacial tension force, the right-hand side consists of the fluid pressure which is given by the osmotic pressure (first term), and of the Maxwell stress due to the tangential electric field (second term). In the construction of the line tension model, we apply the Debye-H\"uckel approximation ($\chi\ll1$). As a consequence, \gleich{eq:TPBsprungdimlos} becomes
\begin{equation}
\frac{h''(\eta)}{\left[1+h'^2(\eta)\right]^{3/2}}=-\psi\left[\phi^2+E_{t}^2 \right]_{(\eta,h(\eta))}.
\label{eq:TDHsprung}
\end{equation}
The quantity~$\psi$ can thus be interpreted as an electrocapillary number measuring the strength of electrostatic forces as compared to interfacial forces. The interfacial shape~$h(\eta)$ influences the electrostatic potential, since the latter is the solution to the problem
\begin{align}
&\Delta\phi=\chi^{-1}\sinh(\chi\phi)\sim\phi~\text{for}~\chi\ll1 \label{eq:DH}\\
&\nabla\phi\cdot\vekn=0~\text{at}~\Sigma_{12}\label{eq:RBowg}\\
&\nabla\phi\cdot\vekn=-1~\text{at}~\Sigma_{13}\label{eq:RBpwg}\\
&\nabla\phi\cdot\vekn\to0~\text{at infinity}\label{eq:RBinf}
\end{align}
In order to find~$h$ and~$\phi$, both of which are required for calculating~$\tau$, we assume a perturbation expansion in powers of~$\psi$ for all dependent variables, e.g.~$\phi=\phiord{0}{}+\psi\phiord{1}{}+\psi^2\phiord{2}{}+\mathcal{O}(\psi^3)$. While incorporating the perturbation expansions into Eqs.~\eqref{eq:DH}, \eqref{eq:RBpwg} and~\eqref{eq:RBinf} is straightforward, condition~\eqref{eq:RBowg} demands careful examination, as it contains the shape~$h(\eta)$ both explicitly and implicitly. In effect, full expansion of the boundary condition~\eqref{eq:RBowg} projects the latter onto the undeformed interface given by~$(\eta,0)$~\cite{vanDyke1964}, as can be seen from the relation
\begin{equation}
\label{eq:entwnachtaylor}
\begin{split}
-\anderstelle{\partiell{\phiord{0}{}}{\zeta}}{\eta,0}+\psi\left[-\anderstelle{\partiell{\phiord{1}{}}{\zeta}}{\eta,0}\right.+\hord{1}{\prime}(\eta)\anderstelle{\partiell{\phiord{0}{}}{\eta}}{\eta,0} \\
-\left.\hord{1}{}(\eta)\left(\phiord{0}{}-\frac{\partial^2\phiord{0}{}}{\partial{\eta}^2}\right)_{\!\!(\eta,0)}\right]
+\mathcal{O}(\psi^2)=0.
\end{split}
\end{equation}
%
\gleich{eq:entwnachtaylor} can be easily rewritten in terms of the polar coordinates shown in Fig.~\ref{fig:Ksys}c. The zeroth-order problem is thus identical to the one treated by D\"orr \& Hardt~\cite{Doerr2012}, allowing to use the model equation $\phiord{0}{}(r,\theta)=\pi/(2\theta)\exp(-c\eta)$ which contains the function~$c=c(\theta)$. Improving upon the empirical expression for~$c$ provided by D\"orr \& Hardt~\cite{Doerr2012}, we present an analytical formula below. By means of the known potential~$\phiord{0}{}$, the exponential first-order shape can be found from \gleich{eq:TDHsprung} as
\begin{equation}
\hord{1}{}(\eta)=Z\left(1-\euler{-2c\eta}\right)
\label{eq:h1}
\end{equation}
with the abbreviation~$Z\definiert\pi^2(1+c^2)/(16c^2\theta^2)$. Insertion of the function~\eqref{eq:h1} into condition~\eqref{eq:entwnachtaylor} and transition to polar coordinates yields the first-order problem 
\begin{align}
&\frac{\partial^2\phiord{1}{}}{\partial r^2}+\frac{1}{r}\partiell{\phiord{1}{}}{r}+\frac{1}{r^2}\frac{\partial^2\phiord{1}{}}{\partial\vartheta^2}=\phiord{1}{}\label{eq:DH1}\\
&\frac{1}{r}\anderstelle{\partiell{\phiord{1}{}}{\vartheta}}{r,\theta}=-\frac{\pi Z}{2\theta}\left[(3c^2-1)\euler{-3cr}+(1-c^2)\euler{-cr}\right]\label{eq:RBowg1}\\
&\frac{1}{r}\anderstelle{\partiell{\phiord{1}{}}{\vartheta}}{r,0}=0,\quad \anderstelle{\partiell{\phiord{1}{}}{r}}{r\to\infty,0\leq\vartheta\leq\theta}=0\label{eq:RBpwg1}
\end{align}
corresponding to the system~\eqref{eq:DH}--\eqref{eq:RBinf}. The solution to Eqs.~\eqref{eq:DH1}--\eqref{eq:RBpwg1} can be found by way of the Kontorovich-Lebedev transform~\cite{Fowkes1998, Chou2001, Kang2003b, Yakubovich1996, Gradshteyn2007}, and reads
\begin{equation}
\phiord{1}{}(r,\vartheta)=\frac{Z}{2\theta}\int_{0}^{\infty}\frac{X(\lambda)\cosh(\lambda\vartheta)}{\sinh(\lambda\theta)} K_{i\lambda}(r)\mathrm{d}\lambda
\label{eq:transrueck1}
\end{equation}
where~$K_{i\lambda}(r)$ is the Macdonald function and
\begin{equation}
\label{eq:Xdef}
\begin{split}
X(\lambda)\definiert{} & i\frac{3c^2-1}{\sqrt{9c^2-1}}\left[\left(3c+\sqrt{9c^2-1}\right)^{i\lambda} -\left(3c+\sqrt{9c^2-1}\right)^{-i\lambda}\right] \\&-i\sqrt{c^2-1}\left[\left(c+\sqrt{c^2-1}\right)^{i\lambda} -\left(c+\sqrt{c^2-1}\right)^{-i\lambda}\right].
\end{split}
\end{equation}
%
%
We are now in a position to find a formal representation of~$\hord{2}{}$ from Eqs.~\eqref{eq:TDHsprung} and~\eqref{eq:transrueck1}. However, as will be demonstrated further below, $\hord{2}{}$ does not influence the second-order line tension and therefore need not be calculated as long as we limit the approximation to second order.
\subsection{Macroscopic and microscopic contact angle}
A first way to exploit the results~\eqref{eq:h1} and~\eqref{eq:transrueck1} is to consider the contact angle. The relation between the microscopic contact angle~$\beta$, for which Young's relation
\begin{equation}
\sigma_{12}\cos\beta=\sigma_{23}-\sigma_{13}
\label{eq:young}
\end{equation}
holds, and the macroscopic contact angle~$\theta$
\begin{equation}
\cos\theta=\cos\beta-\psi\left[1-2\phi(0,\vartheta)\right]~\text{for}~\chi\ll1
\label{eq:thetaexakt}
\end{equation}
was derived by Kang et al.~\cite{Kang2003a}, requiring the potential~$\phi(0,\vartheta)$ at the TCL which they approximated by the zeroth-order potential~$\phiord{0}{}(0,\vartheta)$. The first-order TCL potential
\begin{equation}
\begin{split}
\anderstelle{\phiord{1}{}}{r\to0,\vartheta} =\frac{\pi Z}{2\theta^2}\left[\sqrt{c^2-1} \ln(c+\sqrt{c^2-1})\right.\left.+\frac{1-3c^2}{\sqrt{9c^2-1}}\ln(3c+\sqrt{9c^2-1})\right]\definiertb \frac{\pi Z}{2\theta^2}Y
\label{eq:phi00}
\end{split}
\end{equation}
is given by the distributional limit of \gleich{eq:transrueck1}~\cite{Chou2001}. As a consequence, \gleich{eq:thetaexakt} can now be evaluated up to second order:
\begin{equation}
\cos\theta=\cos\beta+\psi\left(\frac{\pi}{\theta}-1\right)+\psi^2\frac{\pi}{\theta^2}ZY+\mathcal{O}(\psi^3).
\label{eq:thetakang}
\end{equation}
Alternatively, the shape~$\hord{1}{}$ given by \gleich{eq:h1} yields a contact angle relation by itself, whose first-order term must be equal to the one in \gleich{eq:thetakang}. In this way, the analytical expression for the function~$c$ introduced above can be derived, namely
\begin{equation}
\begin{split}
c=\frac{4\theta\left(\pi-\theta\right)}{\pi^2\sin\theta}+\sgn{\!\left(\theta-\frac{\pi}{2}\right)}\sqrt{\left[\frac{4\theta\left(\pi-\theta\right)}{\pi^2\sin\theta}\right]^2-1}.
\end{split}
\label{eq:crel2}
\end{equation}
\subsection{Calculation of the terms contributing to line tension}
As the second consequence of the results~\eqref{eq:h1} and~\eqref{eq:transrueck1}, the line tension~$\tau$ can be determined up to second order. Generally, we consider differences in the total -- electrical plus chemical -- free energy of the system~\cite{Overbeek1990} when deriving~$\dG{1}$ and~$\dG{3}$. The contribution~$\dG{1}$ associated with the shift of the TCL by a distance of~$\dTCL$ (cf. Fig.~\ref{fig:Ksys}a) may be written as
\begin{equation}
\label{eq:dGa1}
\dG{1}=\dTCL\left(\cos\beta+\psi\right)~\text{for}~\chi\ll1.
\end{equation}
The first term in \gleich{eq:dGa1} describes the contribution due to the interfacial energies~$\sigma_{13}$ and~$\sigma_{23}$ and has been reformulated by means of Young's relation~\eqref{eq:young}, while the second term corresponds to the total free energy of a planar EDL of width~$\dTCL$ within the Debye-H\"uckel approximation~\cite{Overbeek1990}. Using Eqs.~\eqref{eq:thetakang} and~\eqref{eq:h1} as well as the relation~$\dTCL=h(\eta\to\infty)\csc\theta$, \gleich{eq:dGa1} can be cast in the form
\begin{equation}
\label{eq:dGa2}
\dG{1}=\dTCL\cos\theta+\psi^2 Z^2\left[\frac{\csc\theta}{Z}-2c\right]+\mathcal{O}(\psi^3).
\end{equation}
The contribution~$\dG{2}$, originating from the change in fluid-fluid interfacial area, can be deduced from Fig.~\ref{fig:Ksys}c and \gleich{eq:h1} and is given by
\begin{align}
\dG{2}&=\int_{0}^{\infty}\!\! \sqrt{1+ h'^2(\eta)}-\sqrt{1+ h'^2(\eta\to\infty)}\,\,\mathrm{d}\eta- L\nonumber\\
&=\frac{1}{2}\psi^2 \int_{0}^{\infty} \hord{1}{\prime 2}(\eta) \,\mathrm{d}\eta+\mathcal{O}(\psi^3)-\dTCL\cos\theta\nonumber\\
&=\frac{c}{2}\psi^2 Z^2+\mathcal{O}(\psi^3)-\dTCL\cos\theta.\label{eq:dGb2}
\end{align}
From Eqs.~\eqref{eq:dGa2} and \eqref{eq:dGb2} we find that the terms $\pm\dTCL\cos\theta$ cancel each other when the sum of all contributions to the line tension is considered. Therefore, as stated above, knowledge of the shape~$\hord{2}{}$ is not required for the second-order line tension. The reason for the terms to drop out is the fact that~$\theta$ is an apparent contact angle obeying the (modified) Young equation~\eqref{eq:thetaexakt} which corresponds to a vanishing free energy variance when the TCL is shifted~\cite{deGennes1985}. Furthermore we note that, since the free energy difference due to the shift of the TCL is already included in~$\dG{1}$, the third contribution~$\dG{3}$ solely consists of the total EDL free energy difference due to deformation of the fluid-fluid-interface at fixed TCL and is thus unaffected by the zeroth-order potential~$\phiord{0}{}$. Hence\cite{Kang2003b,Overbeek1990},
\begin{equation}
\dG{3}=-\psi^2\int_{0}^{\infty}\phiord{1}{}(r,0)\mathrm{d}r+\mathcal{O}(\psi^3)
\label{eq:dGb1formal}
\end{equation}
holds within the Debye-H\"uckel approximation. Furthermore, by insertion of \gleich{eq:transrueck1} and use of the identity~$2\int_0^\infty K_{i\lambda}(r)\mathrm{d}r=\pi\sech(\lambda\pi/2)$ \cite{Gradshteyn2007}, we can write
\begin{equation}
\label{eq:dGb1end}
\dG{3}=-\psi^2\frac{\pi Z}{4\theta} \int_{0}^{\infty}\frac{X(\lambda)}{\sinh(\lambda\theta)\cosh(\lambda\pi/2)}\mathrm{d}\lambda+\mathcal{O}(\psi^3).
\end{equation}
From Eqs.~\eqref{eq:dGa2}, \eqref{eq:dGb2} and~\eqref{eq:dGb1end}, the full line tension is
\begin{equation}
\begin{split}
\tau=\psi^2Z\left[\vphantom{\frac{3}{2}}\csc\theta-\frac{3}{2}cZ\right.\left.-\frac{\pi}{4\theta} \int_{0}^{\infty}\frac{X(\lambda)}{\sinh(\lambda\theta)\cosh(\lambda\pi/2)}\mathrm{d}\lambda\right] +\mathcal{O}(\psi^3)
\end{split}
\label{eq:tau}
\end{equation}
and scales as~$\psi^2$ for small~$\psi$, which is at variance with a result of order~$\psi$ based solely on the electric free energy, as reported by Kang et al.~\cite{Kang2003b}. Here, we consider the total free energy of the microscopic and macroscopic state consistent with the definition of line tension.
\section{Numerical analysis}
\begin{figure}[ht]%
\centering
\subfloat[]{
\includegraphics[width=8.32cm]{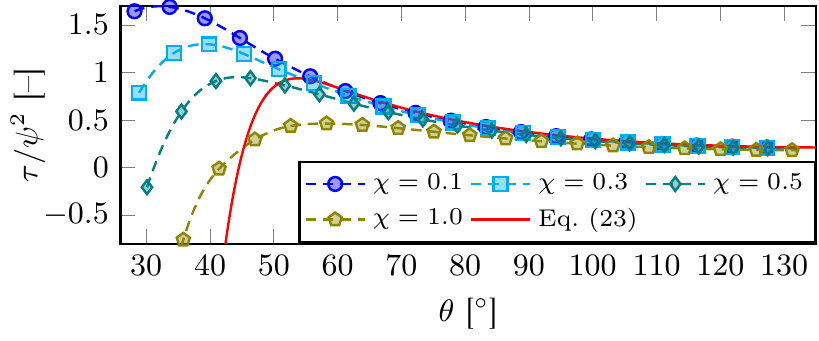}
\label{fig:Pi1tau_Modell}%
}
\\
%
\subfloat[]{
\includegraphics[width=8.32cm]{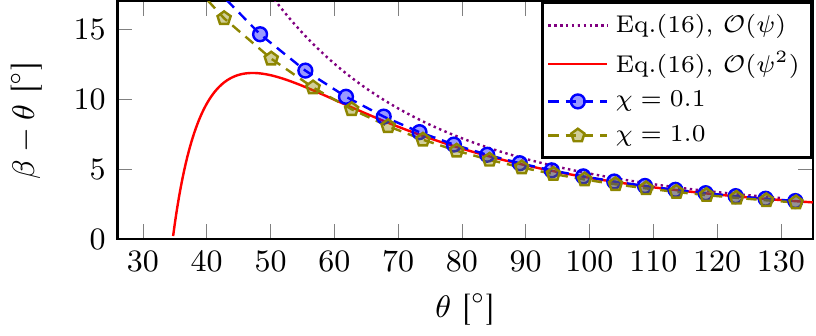}%
\label{fig:Dtheta_Modell}%
}
\caption[Model verification]{Analytical and numerical results for~$\psi=0.1$ and varying~$\chi$: \subref{fig:Pi1tau_Modell} reduced line tension, \subref{fig:Dtheta_Modell} contact angle difference, both as a function of~$\theta$. The lines corresponding to the numerical data represent piecewise cubic spline fits to the data points.}
\label{fig:Modell}
\end{figure}

\begin{figure}[ht]%
\subfloat[]{
\includegraphics[width=8.32cm]{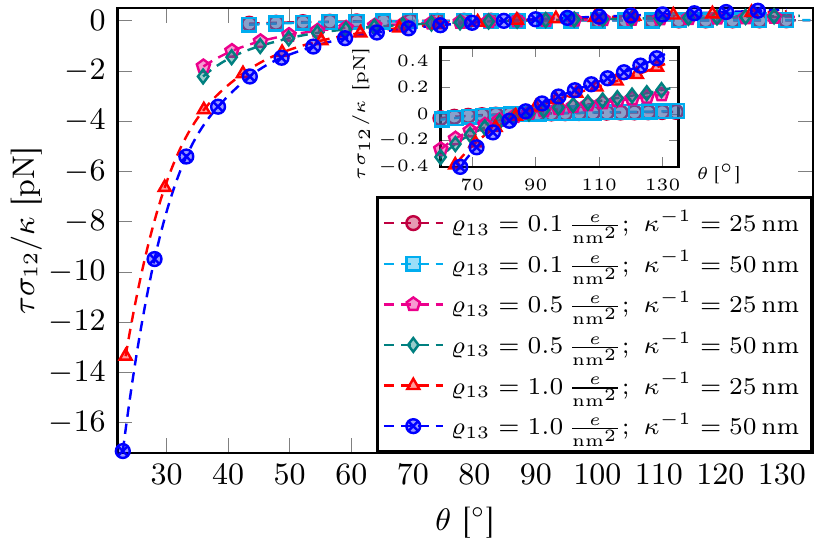}%
\label{fig:tau_COMSOL}%
}
\\
%
\subfloat[]{
\includegraphics[width=8.32cm]{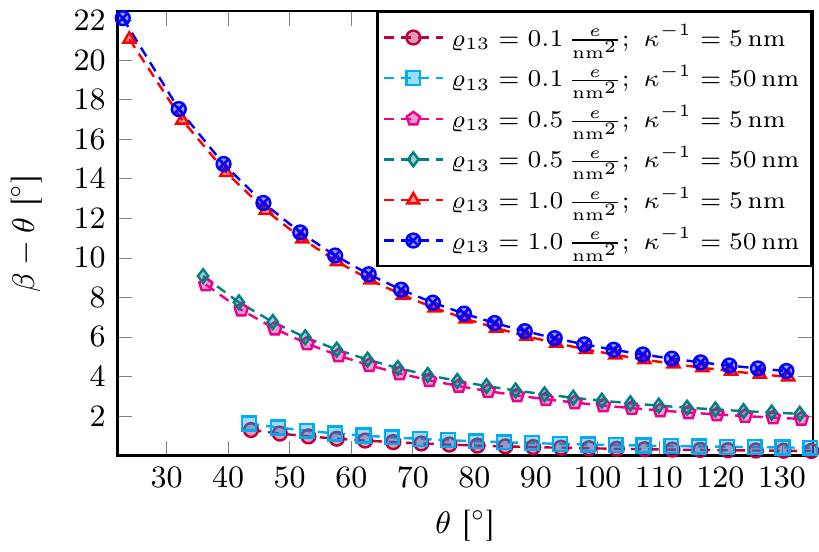}%
\label{fig:Dtheta_COMSOL}%
}
\caption[Nonlinear behaviour]{Numerical results for~$\sigma_{12}=0.072$\,N/m, $\varepsilon_1=78.3\varepsilon_0$, $kT/e=0.0257$\,V: \subref{fig:tau_COMSOL} line tension, \subref{fig:Dtheta_COMSOL} contact angle difference, both as a function of~$\theta$. The lines corresponding to the numerical data represent piecewise cubic spline fits to the data points.}
\label{fig:COMSOL}
\end{figure}

For numerical verification and for studying non-per\-turbative line tension effects, we solved the nonlinear Poisson-Boltzmann problem~\eqref{eq:DH}--\eqref{eq:RBinf} by means of the finite-element solver \textsc{Comsol} Multiphysics\textsuperscript{\textregistered}, starting from a wedge geometry and iteratively adjusting the interfacial shape calculated from \gleich{eq:TPBsprungdimlos}, similar to a procedure in electrowetting studies~\cite{Buehrle2003,Mugele2007}. Accordingly, the appropriate nonlinear expressions for the EDL free energy~\cite{Overbeek1990} were numerically evaluated. Quadratic shape functions and an element size of~$\approx 10^{-3}\kappa^{-1}$ close to the TCL were used. Grid independence of the results was ensured. The model was challenged using a high value of~$\psi=0.1$, which can usually only be achieved in practice by reducing~$\sigma_{12}$ through surfactants. Fig.~\ref{fig:Pi1tau_Modell} demonstrates the good agreement between \gleich{eq:tau} and the numerical calculations up to~$\chi\approx 0.5$ and down to~$\theta\approx 55$\textdegree. Regarding the contact angle difference~$\beta-\theta$ in Fig.~\ref{fig:Dtheta_Modell}, the influence of~$\chi$ is small for the specific choice of~$\psi$. Therefore, the agreement between numerical data and the second-order result~\eqref{eq:thetakang} is reasonable for all values of~$\chi$ shown and for~$\theta$ above~55\textdegree. In order to point out the improvement due to the inclusion of interfacial deformation, the first-order approximation of \gleich{eq:thetakang}, corresponding to the undeformed fluid-fluid interface, is also depicted in Fig.~\ref{fig:Dtheta_Modell}. As can be seen from Figs.~\ref{fig:Pi1tau_Modell} and~\ref{fig:Dtheta_Modell}, the model is not suited to represent the region of~$\theta<55$\textdegree. That fact must be attributed to the potential distribution which cannot be accurately represented within the assumptions made by D\"orr \& Hardt\cite{Doerr2012}, namely an exponential potential distribution at the fluid-fluid interface, regardless of the functional form of~$c$. In particular, the contact angle difference mainly depends on the potential value at the TCL (cf. \gleich{eq:thetakang}) and is thus directly affected by the accuracy of the model for the potential distribution. Therefore, the model can be considered valid only above~$\theta\approx55$\textdegree{}.
Regarding the nonlinear Poisson-Boltzmann regime, it is instructive to consider a realistic system where the fluids are water and air, respectively. According to Fig.~\ref{fig:tau_COMSOL}, $\tau$~decreases towards large negative values when~$\theta$ is small, an effect pronounced at large interfacial charge density~$\varrho_{13}$.
For the latter, high but physically reasonable values were chosen~\cite{Lyklema1995,Frydel2011}, given in the legend of Figs.~\ref{fig:tau_COMSOL} and~\ref{fig:Dtheta_COMSOL}. A sign change of~$\tau$ with varying~$\theta$ is evident from Figs.~\ref{fig:Pi1tau_Modell} and~\ref{fig:tau_COMSOL} (cf. the magnified region within Fig.~\ref{fig:tau_COMSOL}). The difference between~$\beta$ and~$\theta$, depicted in Fig.~\ref{fig:Dtheta_COMSOL}, is mainly governed by~$\varrho_{13}$ and only slightly affected by the Debye length.
\clearpage
\section{Summary and conclusions}
In summary, to the best of our knowledge we performed the first analytical and numerical analysis of electrostatic line tension based on the commonly accepted picture of "microscopic" and "macroscopic" states in wetting scenarios. Agreement between our model and the numerical results over a significant range of the chosen perturbation parameter was demonstrated. As a striking property of the electrostatic line tension we found a change in sign from positive to negative when the macroscopic contact angle~$\theta$ is decreased. With~$\theta$ approaching zero, the line tension apparently diverges as a result of the growing discrepancy between the micro- and macroscopic state. For~$\theta \approx 20$\textdegree, for example, the electrostatic line tension of water can assume values of around~-15\,pN (cf. Fig.~\ref{fig:tau_COMSOL}), similar to experimental line tension values recently reported, such as -30\,pN~\cite{Guillemot2012} and~-35\,pN~\cite{Heim2013}. Therefore, the electrostatic line tension in situations of high interfacial charge density and small contact angle may significantly contribute to experimentally determined line tension values. Since the electrostatic line tension is mainly governed by the interfacial charge density, it is tunable via the pH value of the electrolyte. This mechanism could be utilized in context with nanoparticles adsorbed at fluid-fluid interfaces, in the sense that negative line tension stabilizes the particles as opposed to positive line tension. Accordingly, the stability of interfacial nanoparticles is expected to be a function of the pH value of the electrolyte phase. As a possible extension of the theory presented here, the electrostatic line tension in the presence of surfactants is expected to attain significant values and thus deserves attention.
\begin{acknowledgments}
We gratefully acknowledge financial support by the German Research
Foundation through grant number HA 2696/25-1. 
\end{acknowledgments}

\end{document}